\documentclass[runningheads]{svmult}
\usepackage{makeidx}   
\usepackage{graphicx}  
\usepackage{subeqnar}  
\usepackage{multicol}  
\usepackage{cropmark} 
\usepackage{physprbb}  
\newcommand{\eqb}{\begin{equation}}
\newcommand{\eqe}{\end{equation}}
\newcommand{\dmb}{\begin{displaymath}}
\newcommand{\dme}{\end{displaymath}}
\newcommand{\pd}{\partial}
\newcommand{\ep}{\varepsilon}
\newcommand{\eab}{\begin{eqnarray}}
\newcommand{\eae}{\end{eqnarray}}

\newcommand{\e}{\mbox{e}}
\newcommand{\be}{\begin{equation}}
\newcommand{\ee}{\end{equation}}

\newcommand{\La}{\Lambda}

\begin{document}
\title*{Gauged Inflation}
\toctitle{Gauged Inflation} 
\author{Ralf Hofmann and Mathias Th. Keil}
\institute{Max-Planck-Institut f\"ur Physik\\ 
Werner-Heisenberg-Institut\\ 
F\"ohringer Ring 6, 80805 M\"unchen\\ 
Germany}

\maketitle              

\begin{abstract}
We propose a model for cosmic inflation 
which is based on an effective description of strongly 
interacting, nonsupersymmetric matter within 
the framework of dynamical abelian projection and centerization. 
The underlying gauge symmetry is 
assumed to be SU(N) with N$\gg$1. Appealing to a 
thermodynamical treatment, the ground-state structure 
of the model is determined by a potential for the inflaton field (monopole condensate) which allows 
for nontrivially BPS saturated and thereby stable solutions. For $T<M_P$ this leads 
to an apparent decoupling of 
gravity from the inflaton dynamics. The ground state dynamics implies a 
heat capacity for the vacuum leading to inflation for 
temperatures comparable to the mass scale $M$ of the potential. 
The dynamics has an attractor property. In contrast to the usual 
slow-roll paradigm we have $m\gg H$ during inflation. As a consequence, density perturbations 
generated from the inflaton 
are irrelevant for the formation of large-scale structure, and the model has 
to be supplemented with an inflaton independent mechanism for the generation of 
spatial curvature perturbations. 
Within a small fraction of the Hubble time 
inflation is terminated by a 
transition of the theory to its center symmetric phase. Due to the prevailing 
$Z_{\tiny\mbox N}$ symmetry relic vector bosons are stabilized and therefore potential 
originators of UHECR's beyond the GZK bound.   
 
\end{abstract}

\section{Introduction}

Cosmic Inflation \cite{Guth,Linde0,Linde1} is the concept 
to explain the almost perfect isotropy, 
homogeneity, and flatness of the observable 
universe on large scales and the fact that no topological defects of GUT 
phase transitions have been observed. To describe this phenomenon 
in the framework of Friedmann-Robertson-Walker (FRW) geometry 
one usually assumes cosmology to be driven by one or more real and minimally coupled
scalar fields which roll down their respective potentials. If a regime of slow-roll is reached, 
where the energy density is sufficiently dominated by the
potential, the universe undergoes a rapid expansion stretching a 
causally connected patch to scales much larger than the horizon. 
As a result, after inflation the observable universe remains highly 
homogeneous and isotropic as long as regions, which were causally disconnected prior to inflaton, 
do not enter the horizon. The measured homogeneity and isotropy of the observable universe 
then implies that inflation must have produced 
at least 60 e-foldings of the scale factor.

Within the above approach the slow-roll paradigm is exploited to explain 
the small deviations from perfect isotropy in the 
cosmic microwave background (CMB) and also the formation of large-scale structure. The idea is that 
quantum fluctuations of the inflaton field are so strongly red-shifted during 
inflation that their wavelengths may leave the horizon. On super-horizon scales these fluctuations transform into 
classical gaussian noise with a scale invariant spectrum. Upon horizon entry sufficiently long 
after reheating these classical perturbations are the originators of spatial curvature perturbations which 
cause the formation of large-scale structure \cite{Lythbook}. Crucial for the validity of this picture 
is the validity of the condition $m\ll H$ following from the slow-roll paradigm. Here $m$ is the mass of 
fluctuations, and $H$ is the Hubble parameter during inflation. Explaining perturbations by the same field 
that causes inflation is a rather economic feature. However, $m\ll H$ seems 
rather artificial since it expresses a hierarchy in the mass scales governing the 
{\sl matter} dynamics which drives inflation. 

Since inflationary models are not very constrained progress is achieved by relying on a strong 
principle which determines the dynamics and at the same time {\sl explains} the usual, 
phenomenologically driven ad hoc assumptions and desired features. 
For example, if inflation can be made responsible for the generation 
of some of the massive and (quasi)stable relics which constitute cold dark matter and upon decay 
may cause ultra-high energy cosmic rays (UHECR) beyond 
the Greisen-Zatsepin-Kuzmin (GZK) cutoff \cite{Rubakov} 
then it should also explain {\sl why}. Or, {\sl why} does the inflaton potential not allow for moduli whose 
occurence after inflation can cause large inhomogeneities and anisotropies by 
isothermal perturbations \cite{Randall,Freese} being in contradiction with observation? 
Another question is whether the slow roll of the inflaton field during inflation rather than be imposed 
can be explained.

In this talk I propose inflation to be caused by the dynamics of 
strongly interacting matter which is governed by a nonabelian 
{\sl gauge} symmetry. For definiteness we 
assume this symmetry to be SU(N). Within an effective, thermal description, which is based on 
dynamical abelian projection at high and centerization at low energy, we are able to positively address the 
points raised above.

\section{The model}

\subsection{Effective description of thermalized pure SU(N) gauge theory}

We consider FRW geometry. The matter sector is assumed to be effectively described, minimally coupled, and 
gauged pure SU(N) thermodynamics. We appeal to the concept of dynamical abelian projection 
\cite{thooft,dualsc} and centerization \cite{centerlat} 
of the fundamental gauge symmetry. In this approach, N--1 species of 
condensed magnetic monopoles give mass to N--1 species of 
abelian gauge bosons at temperatures $T$ larger than a dynamically 
generated mass scale $M$. For $T\sim M$  the maximally 
abelian gauge symmetry U(1)$^{\tiny{\mbox{N}}}$ reduces to the discrete center symmetry $Z_{\tiny\mbox N}$ so 
that gauge bosons do not participate in the dynamics anymore. Since it seems to be 
impossible to apply the framework of ref.\,\cite{dualsc} to the case of 
arbitrarily large N we simplify the picture by assuming a single complex U(1) gauge symmetry, 
but we keep the feature of reduction to the center symmetry for $T\sim M$. 
The action then assumes the following form
\eqb
\label{act}
S=\int d^4x \sqrt{-g}\left[
- \frac{1}{16\pi G}R - \frac{1}{4} F_{\mu\nu}F^{\mu\nu} + \overline{{\cal D}_\mu \phi}{\cal D}^\mu \phi - 
V(\bar\phi \phi)\right]\, ,
\eqe
where ${\cal D}_\mu\equiv\pd_\mu+ieA_\mu$ denotes the gauge 
covariant derivative, and $V$ is an effective potential 
to be constructed below. A thermal treatment needs a continuation of (\ref{act}) 
to euclidean signature with a compact time coordinate $0\le\tau\le\beta\equiv 1/T$ along  
which the inflaton field $\phi$ solely varies. Speaking of analytical continuation, 
let me stress at this point the fact that observables 
such as energy-density or the Hubble parameter being 
time-independent in the euclidean have a trivial continuation to 
Minkowskian signature. 

We now construct the potential $V$. Disregarding gravity and the gauge sector for the moment, 
we want $\phi$, $\bar{\phi}$ to be stable solutions to the scalar field equations. This criterion is automatically satisfied 
if we assume $\phi$, $\bar{\phi}$ to be BPS saturated \cite{BPS} since it is this class of 
field configurations possessing lowest euclidean action in a given topological sector. 
As we will show later, a consequence of BPS saturation is an efficient decoupling of gravity away from 
the Planck scale. On the other hand, for a global, thermal 
description of the ground state of the system we want only a single, dynamically
generated mass scale $M$ to appear in the potential. 
According to the above picture of successive symmetry reduction 
the potential is gauge invariant for $T>M$ and $Z_{\tiny\mbox N}$ symmetric for $T\sim M$\footnote{
$T>M$ and $T\sim M$ are merely labels and not to be taken literally. 
For example, letting $N\to\infty$ we have $T>M/2\pi$ in the regime of continuous
gauge symmetry.}. It will be shown below that this 
implies the validity of a Born-Oppenheimer approximation in the regime $T>M$ such that the back-reaction 
gauge field onto the scalar dynamics can be neglected. 

The above constraints determine the potential uniquely \cite{Losev}, and we have 
\eab
\label{pot}
V(\bar{\phi}\phi)&\equiv&{\bar{V}}^{1/2}{V}^{1/2}=\frac{M^6}
{\bar{\phi}\phi}+\lambda^2 M^{-2(N-3)}(\bar{\phi}\phi)^N-2\,\lambda M^{6-N}\frac{1}
{\bar{\phi}\phi}\mbox{Re}\,\phi^{N}\ ,\ \ \ \mbox{or}\nonumber\\ 
{V}^{1/2}&=&\frac{M^3}{\phi}-
\lambda \frac{\phi^{N-1}}{M^{N-3}}\ ,\ \ \ \lambda\sim 1\,.
\eae
The corresponding BPS equations read 
\eqb 
\label{BPS}
\pd_\tau\phi=\bar{V}^{1/2}\,,\ \ \ \ \pd_\tau\bar{\phi}={V}^{1/2}\,.
\eqe
Note that adding a {\sl constant} to the potential would 
destroy the feature that BPS saturated solutions exist.

\subsection{Ground state solution: BPS saturated thermodynamics}  

I now come to the solution of the ground state dynamics for $T>M$, 
that is, the regime of continuous gauge symmetry where $V=M^6/\bar{\phi}\phi$. 
Obviously, $\bar{V}^{1/2}$ and ${V}^{1/2}$ are only determined up to phases $\e^{i\delta}$ and 
$\e^{-i\delta}$, respectively, which at first sight 
expresses the freedom to choose the gauge in which one would like to solve (\ref{BPS}). 
At finite temperature and for a bosonic field 
the physical gauge is the one that yields {\sl periodic} solutions\footnote{By physical gauge we mean 
the one that maximizes the physics contained in the scalar sector alone. 
To determine this gauge, the periodicity criterion is applied.}. Relying on the definition (\ref{pot}), 
it turns out that only for the choice $\delta=\pm \pi/2$ 
there are periodic solutions to (\ref{BPS}) which read
\eqb
\label{BPSsol}
\phi^{(n)}(\tau)=\sqrt{\frac{M^3\beta}{2|n|\pi}}\,
\e^{2n\pi i\frac{\tau}{\beta}}\ ,\ \ (n\in{\bf Z})\ .
\eqe
Different $n$ label different topologies. We restrict ourselves to $n=1$ in the following. 
There are two important observations. First, calculating the ratio of the masses of 
scalar and vector excitations in the background of (\ref{BPSsol}), we obtain
$m_{\phi}/m_A\ge\sqrt{6}/{e}$. Hence, the scalar field is slow as compared to 
the gauge field if $e$ is not larger than unity, and therefore the Born-Oppenheimer approximation 
leading to (\ref{BPSsol}) is justified. Second, the ratio of scalar mass to temperature is 
$m_\phi/T=\sqrt{6}\times 2\pi\sim 15.4$. This means that $V$ indeed is an 
{\sl effective} potential since fluctuations in $\phi$ are irrelevant and therefore are 
contained in the shape of $V$ and in the BPS saturated solution $\phi^{(1)}$.

Let us now look at the gauge field dynamics of the ground state. We have to solve 
\eqb
\label{ME}
\pd_\mu\left[\,\sqrt{\tilde{g}} F_{\mu\nu}\,\right]=j_\nu\ ,
\eqe
where $j_\mu\equiv ie\,\delta_{\mu 0}\ [{\bar\phi}^{(1)}\stackrel{\rightarrow}{\cal D}_\tau \phi^{(1)}-
(\overline{\phi^{(1)}\stackrel{\leftarrow}{\cal D}_\tau})\phi^{(1)}]$. 
If $A_\mu$ is pure gauge\footnote{Ground state is assumed to be locally Lorentz invariant.} 
(\ref{ME}) implies that ${\cal D}_\tau \phi^{(1)}$ vanishes. Accordingly, the solution to the gauge field dynamics 
boils down to finding an $A_\mu=\delta_{\mu 0}A(\tau)$ such that ${\cal D}_\tau \phi^{(1)}=0$. This is most easily done 
in the unitary gauge $\phi^{(1)}=|\phi^{(1)}|$ where due to (\ref{BPSsol}) 
we have $\pd_\tau\phi^{(1)}=0$. Hence, ${\cal D}_\tau \phi^{(1)}=0$ for $A(\tau)\equiv 0$ 
in this gauge. 
  
Inserting the solution to the ground state dynamics into the matter sector 
of the euclidean action, the only nonvanishing 
term is the potential with 
\eqb
\label{La}
V(|\phi^1|^2)=2\pi M^3 T\ .
\eqe
This does not depend on euclidean time and therefore is 
interpreted as a $T$ dependent 
cosmological constant $\La=\La(T)$ in Minkowskian signature.

\subsection{Excitations}  

We already realized that there are no scalar excitations for $T>M$. 
The excitations of the gauge field 
are massive ($m_A=e|\phi^{(1)}|$) due to the Higgs mechanism, 
and therefore we have three polarization states. We assume that the gauge coupling $e$ 
is small enough to permit a calculation of the energy density and the pressure based on a perturbation of 
the black-body expression. We obtain 
\eqb
\label{epT}
\ep_A=\frac{\pi^2}{10}T^4-\frac{e^2}{16\pi} M^3 T\ ,\ \ \ \ 
p_A=\frac{\pi^2}{30}T^4-\frac{e^2}{24\pi} M^3 T\,. 
\eqe
For example, taking $e\sim0.1$, we would trust (\ref{epT}) since $M\le 2\pi T$.
 
\subsection{Energy-momentum conservation in a Friedmann universe}

Taking into account the equation of state $p_\La=-\ep_\La$ for the vacuum, 
energy and pressure of the universe are given as   
\eqb
\label{pf}
\ep=\ep_A+\La\ ,\ \ \ \ \ p=p_A-\La\ .
\eqe
In terms of $\La$ this reads
\eqb
\label{epL}
\ep=\frac{\Lambda^4}{160\pi^2 M^{12}}+\kappa_1 \Lambda\ ,\ \ \ \ 
p=\frac{\Lambda^4}{480\pi^2 M^{12}}-\kappa_2 \Lambda\ , 
\eqe
where
\eqb
\label{kap}
\kappa_1\equiv 1-\frac{e^2}{32\pi^2}\ ,\ \ \ \kappa_2\equiv 1+\frac{e^2}{48\pi^2}\ .
\eqe
Energy-momentum is conserved if the scale factor $a$ satisfies 
\eqb
\label{FrieI}
\frac{d}{da} \left(\ep\,a^3\right)=-3\,a^2\, p\ .
\eqe
Upon use of (\ref{epL}) eq.\,(\ref{FrieI}) yields 
\eqb
\label{Lp}
\frac{d}{da}\Lambda=-\frac{\Lambda}{a}\,
\frac{\frac{\Lambda^3}{40\pi^2M^{12}}+3(\kappa_1-\kappa_2)}{\frac{\Lambda^3}{40\pi^2M^{12}}+
\kappa_1}\ .
\eqe

\subsection{Attractor property of inflaton dynamics}

The solution to eq.\,(\ref{Lp}) can be given analytically only 
for the inverse function $a/a_0=(a/a_0)(\Lambda)$. We have
\eqb
\label{sol}
\frac{a}{a_0}=\left(\frac{\Lambda}{\Lambda_0}\right)
^{\frac{\kappa_1}{3(\kappa_2-\kappa_1)}}\left(\frac{\Lambda^3+120\pi^2(\kappa_1-\kappa_2)M^{12}}
{\Lambda_0^3+120\pi^2(\kappa_1-\kappa_2)M^{12}}\right)
^{\frac{2\kappa_1-3\kappa_2}{9(\kappa_2-\kappa_1)}}\ .
\eqe
Using (\ref{epL}) and (\ref{sol}) it can be shown that for $\La\gg M^4$ 
there is radiation scaling, $\ep\sim a^{-4}$, whereas for $\La\sim
M^4$ we have $\ep\sim\mbox{const}$. The points $\La/M^4$, where $a/a_0$ has undergone 60 e-foldings, 
turn out to be between 1.34 and 1.44 for 
the three very different initial conditions $\La_0/M^4=10^2,10^4,10^7$! So the (inflationary) regime, 
where the cosmological constant dominates the energy density, does 
practically not depend on the initial conditions prescribed at the 
borderline of applicability of our model. Hence, we demand that inflation 
be terminated at $\La/M^4=1.4$ by a transition to the center symmetric phase. This phase transition sets in at the first 
point of inflexion $|\phi_c|$ of $V$. Setting $V(|\phi_c|^2)=\La=1.4\,M^4$, yields N=34. 
We will work with this value in the following.

\subsection{Decoupling of gravity from inflaton dynamics}

Let me now come back to the question how well the effect of gravitational back-reaction 
onto the inflaton dynamics is already contained in the shape of the 
potential considering the BPS saturated solution $\phi^{(1)}$. 
To decide on this we look at the consequences of this BPS saturation. 
Taking the covariant divergence of eqs.\,(\ref{BPS}) and appealing again to (\ref{BPS}), 
we obtain the following right-hand sides
\eab
\label{RHS} 
\pd_\tau\bar{V}^{1/2}+3H\bar{V}^{1/2}&=&
\frac{\pd \bar{V}^{1/2}}{\pd \bar{\phi}}{V}^{1/2}+3H\bar{V}^{1/2}=
V^{-1/2}\frac{\pd V}{\pd \bar{\phi}} V^{1/2}+3H\bar{V}^{1/2}\nonumber\\ 
&=&\frac{\pd V}{\pd \bar{\phi}}+3H\bar{V}^{1/2}\,,\nonumber\\ 
\pd_\tau V^{1/2}+3H{V}^{1/2}&=&
\frac{\pd {V}^{1/2}}{\pd {\phi}}\bar{V}^{1/2}+3H{V}^{1/2}= 
\bar{V}^{-1/2}\frac{\pd V}{\pd \phi} \bar{V}^{1/2}+3H{V}^{1/2}\nonumber\\ 
&=&\frac{\pd V}{\pd \phi}+3H{V}^{1/2}\,.
\eae
The terms $\frac{\pd V}{\pd \bar{\phi}}$ and $\frac{\pd V}{\pd \phi}$ 
in the respective right-hand sides for $\bar{\phi}$ and $\phi$ represent 
the usual sources in the 
scalar equations of motion. The terms $3H\bar{V}^{1/2}$ 
and $3H{V}^{1/2}$ are in excess. Therefore, we must investigate whether 
the ratio 
\eqb
\label{R}
R\equiv\left|\frac{3H\,V^{1/2}}{\frac{\pd V}{\pd \phi}}\right|
\eqe
is smaller than unity. For the radiation dominated epoch and during inflation we respectively obtain
\eqb
R_r\sim\frac{3\sqrt{\frac{8}{5}\pi^3 M^3T}\,T^2/M_P}{(2\pi\,M\,T)^{3/2}}=\sqrt{9/5}\frac{T}{M_P}\,, \ \ \ 
R_i\sim 2\sqrt{18}\frac{M}{M_P}<10^{-3}\ 
\eqe
for $M$ being smaller than the GUT scale $\sim 10^{16}$ GeV. 
If $T$ is of order $M_P$ $R_r$ is of order unity. So decoupling of gravity becomes effective 
if the initial temperature is smaller than the Planck mass. Since there is an attractor property of 
the dynamics in the sub-Planckian regime we have to assume that Planckian 
physics drives the universe towards temperatures lower than $M_P$ to be able to apply 
our model. Summarizing, we have shown that back-reacting gravity is contained in the 
BPS saturated scalar dynamics if initial 
conditions are set below the Planckian regime.

\section{Numerical solutions}

\subsection{Flat universe}

Let us for definiteness assume $M=10^{13}\,$GeV 
and Planckian energy density of the initially dominating radiation and 
solve the Friedmann equation 
\eqb
\label{FrieII}
H^2(t)\equiv\left(\frac{\dot{a}(t)}{a(t)}\right)^2=\frac{8\pi G}{3}\left(\frac{\Lambda^4}{160\pi^2 M^{12}}+
\kappa_1 \Lambda-\frac{3k}{8\pi G a^2(t)}\right)\, 
\eqe
numerically for the cases of closed, flat, and open universes corresponding to 
$k=+1,0,-1$, repectively. The results are shown in Fig.\,1. 
\begin{figure}
\vspace{4.5cm}
\includegraphics{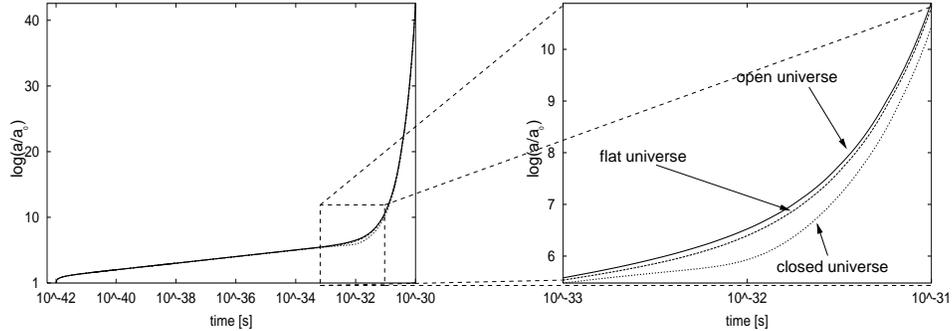}
\caption{Evolution of open, flat and closed universes. We have set 
$a_{0,r}=1/8.65\times10^{6}M_P^{-1}$ which is very near to the 
critical radius for a closed universe.} 
\end{figure}

\subsection{Closed universes need not collapse.}

Let me now raise the question under which conditions a closed universe of Planckian initial condition 
can escape gravitational 
collapse. To answer this we need to investigate whether the Hubble parameter   
\eqb
\label{Hubble}
H=\sqrt{\frac{8\pi G}{3}\left(\frac{\Lambda^4}{160\pi^2 M^{12}}+
\kappa_1 \Lambda-\frac{3}{8\pi G a^2(t)}\right)}\  
\eqe
can be prevented from becoming zero for suitably chosen 
initial radii $a_0$. As a consequence of the finite heat capacity of the vacuum and with $M=10^{13}\,$GeV 
it turns out that for 
\eqb
\label{ab}
a_{0}>\frac{1}{7.3}\times 10^6 M_P^{-1}\sim 10^{5} M_P^{-1}\ 
\eqe
$H$ always remains positive, and therefore the corresponding 
closed universe does not collapse. However, eq.\,(\ref{ab}) 
expresses a hierarchy originated by a hierarchy in the 
scales of gravity and matter. This stands as a fact as long as we have no unified quantum 
description of gravity and matter at our disposal to decide 
whether such a hierarchy can be dynamically generated or not.
 
\section{Universe after inflation}

\subsection{Termination of inflation and reheating}

In this last part of my talk I will briefly mention what our model has to say concerning the epochs following 
quasiexponential expansion. A termination of inflation is induced by the transition of the SU(N) gauge theory from 
its phase of maximally abelian dynamics to that governed by the 
discrete center symmetry $Z_{\tiny\mbox N}$. In our model this happens 
around the point of inflexion $|\phi_c|$ where inflaton fluctuations become massless and therefore large. Even 
solutions to the euclidean mean field equations yield a time dependent inflaton amplitude indicating the 
breakdown of thermal equilibrium in this regime. Therefore, we have to solve 
the dynamics in {\sl Minkowskian} signature. 
Fluctuations around the mean field grow like $\delta\phi \sim e^{mt}$ for tachyonic masses. 
Numerically, $m$ is of order $M$. On the other hand, the Hubble parameter $H$, which can be taken inflation-valued, 
is $H\sim\frac{M}{M_P}M\sim 10^{-6} M$. This implies a duration of the 
tachyonic regime of about $10^{-6}$ inflationary Hubble times. Moreover, during 
the subsequent regime of reheating (positive mass squared for the fluctuations) 
$\phi$ performs damped oscillations about its vacuum value $\phi=M$. 
If there were only a few oscillations then this regime would last 
again only about $10^{-6}$ Hubble times since the frequency of oscillation should be comparable to the 
mass of excitations $=NM=34\,M$ \cite{hof5}. 
There seems to be numerical evidence that the assumption of only a few oscillations is justified 
\cite{Bellido,KofmanLinde,Parry1}. Therefore, inflation would be terminated very rapidly in our scenario.  

\subsection{Adiabatic density perturbations due to the inflaton field?}

Usually, one estimates the magnitude of 
adiabatic density perturbations stemming 
from the inflaton field by 
exploiting $m\ll H$ during inflation which is a 
consequence of the slow-roll paradigm for the dynamics in 
Minkowskian signature. In our model this condition is 
not met for the bulk of inflation. On the contrary, 
we have $m\sim M_P/M H$. Thanks to the heat capacity of the vacuum 
thermal equilibrium is intact during the bulk of inflation and only gets destroyed 
with the onset of the phase transition. Therefore, we may estimate the size of fluctuations by looking at 
the occupation numbers of scalar excitations being of the order $10^{-7}$
\footnote{The energy density of vector excitations is suppressed by 
a factor $(2\pi)^{-4}$, and hence we assume their fluctuations too be negligible.}. 
Hence, there are no fluctuations during the bulk of inflation. 
A rough estimate for the density contrast arising from 
the few excited scalar modes that may leave the 
horizon at a very late stage of inflation and 
are assumed to enter it as classical perturbations well after reheating yields
\eqb
\label{dens}
\frac{\delta T}{T}\sim\frac{\delta\rho}{\rho}\equiv\left.
\frac{\pd V/\pd\phi}{V}\,\delta{\phi}\right|_{\phi=\phi_c}<10^{-9}\ .
\eqe
This is much too low to explain the measured anisotropy $\frac{\delta T}{T}\sim 10^{-5}$ 
of the CMB. Therefore, the required density perturbations do not 
originate from the fluctuations of the 
inflaton field in our model. Recently, it was proposed by Lyth and Wands \cite{Lyth} 
that the spatial curvature perturbations required for the formation 
of the large scale structure can originate from an extra 
scalar field {\sl not} driving inflation -- the curvaton. Our model has to be supplemented with 
such a mechanism for the generation of density perturbations.

There are two final points which I would like to stress. First, gauged inflation excludes the post-inflationary 
occurence of moduli. Moduli fluctuations are associated with a mass scale much smaller 
than the one of the field driving and terminating inflation. Therefore, in contradiction to observation, 
they have the potential to introduce large isothermal 
density perturbations. Typical candidates are the Goldstone modes of the spontaneous 
breakdown of a {\sl continuous, global} symmetry \cite{Freese}. 
In our model we have a spontaneous breakdown of a $Z_{\tiny\mbox N}$ symmetry during tachyonic preheating. 
However, (a) this symmetry originated from a gauge symmetry \cite{Wilczek}, and (b), 
it is not {\sl continuous}. Second, the vector bosons surviving inflation are 
coupled in a $Z_{\tiny\mbox N}$ symmetric way to the matter generated during 
reheating. Thus, their decay is mediated by high 
dimensional operators which makes them quasistable and therefore candidates 
for the originators of UHECR's beyond the GZK bound and cold dark matter \cite{Cohn}.

\section{Summary}

In this talk I presented a model for cosmic inflation which is based on the 
effective, thermal description of an SU(N) pure gauge theory at large N and FRW gravity. As a consequence 
of the BPS saturation of the ground state, which is a dynamical 
monopole condensate, gravitational and gauge field dynamics do not back-react. 
Cosmic evolution is driven by a competition between the finite heat capacity of the 
vacuum and the radiation of massive gauge field quanta. This leads to inflation for 
temperatures comparable to the mass scale of the gauge 
theory within large variations of the initial conditions. 
A closed universe of Planckian initial conditions need not collapse if its initial radius is larger than a critical value determined by the hierarchy between $M_P$ and the 
scale of the matter sector. In contrast to the usual slow-roll paradigm 
inflation is realized at $m\gg H$ {\sl although} the 
time variation of the inflaton field is nil. Density 
perturbations stemming from adiabatic inflaton 
fluctuations during inflation are irrelevant for the formation of large-scale 
structure and CMB anisotropy. Hence, 
the model needs to be supplemented with a mechanism to 
generate them. This mechanism is available \cite{Lyth}. 

Inflation is very rapidly terminated in the course of a 
transition from the phase of maximal abelian gauge symmetry to 
the phase of discrete center symmetry. Due to the latter 
there are no moduli excitations which possibly could introduce 
large isothermal density perturbations after inflation. Inflationary relics are massive vector 
bosons which are protected by high dimensional operators from 
decaying into matter generated during reheating. Therefore, these vector particles 
may originate UHECR's beyond the GZK bound. On the other hand, 
they do contribute to the cold dark matter 
of the universe. 

\section*{Acknowledgements}

I would like to thank the organizers for providing the framework for a very stimulating 
conference.  

\bibliographystyle{prsty}

\begin{thebibliography}{10}      
 
\bibitem{Guth}
A.H. Guth, Phys. Rev. {\bf D23}, 347 (1981).

\bibitem{Linde0}
A.D. Linde, Phys. Lett. {\bf B108}, 389 (1982).\\ 
A. Albrecht and P.J. Steinhardt, Phys. Rev. Lett. {\bf 48}, 1220 (1982).\\ 
A.H. Guth and P.J. Steinhardt, Sci. Am., May 1984, p. 90.

\bibitem{Linde1}
A.D. Linde, Rep. Prog. Phys. {\bf 47}, 925 (1984).

\bibitem{Lythbook}
A.R Liddle and D.H. Lyth, {\sl Cosmological Inflation and Large Scale Structure}, Cambridge University Press (2000). 

\bibitem{Rubakov}
V. Berezinsky, M. Kachelriess, A. Vilenkin, Phys. Rev. Lett. {\bf 79}, 4302 (1997), astro-ph/9708217.\\ 
V.A. Kuzmin and V.A. Rubakov, Phys. Atom. Nucl. {\bf 61}, 1028 (1998), astro-ph/9709187.

\bibitem{Randall}
L. Randall and S. Thomas, Nucl. Phys. {\bf B449}, 229 (1995), hep-ph/9407248.

\bibitem{Freese}
M. Bucher and Y. Zhu, Phys. Rev. {\bf D55}, 7415 (1997), astro-ph/9610223.\\ 
A. Dolgov and K. Freese, Phys. Rev. {\bf D51}, 2693 (1995), hep-ph/9410346.\\ 
K. Freese, Phys. Rev. {\bf D50}, 7731 (1994), astro-ph/9405045. 

\bibitem{thooft}
G. 't Hooft, Nucl. Phys. {\bf B190}, 455 (1981).

\bibitem{dualsc}
T. Suzuki, Prog. Theor. Phys. {\bf 80}, 929 (1988); {\bf 81}, 752 (1989).\\ 
S. Maedan and T. Suzuki, Prog. Theor. Phys. {\bf 81}, 229 (1989).\\ 
H. Ichie, H. Suganuma and H. Toki, Phys. Rev. {\bf D 52}, 2944 (1995).\\ 
M.N. Chernodub, F.V. Gubarev, M.I. Polikarpov, 
V.I. Zakharov, Nucl. Phys. {\bf B600},163 (2001), hep-th/0010265. 

\bibitem{centerlat}
L. Del Debbio, M. Faber, J. Greensite, and S. Olejnik, Phys. Rev. {\bf D55}, 2298 (1997), hep-lat/9610005.\\ 
P. de Forcrand and M. D'Elia, Phys. Rev. Lett. {\bf 82}, 4582 (1999), hep-lat/9901020.

\bibitem{BPS}
E.B. Bogomolny, Sov. J. Nucl. Phys. {\bf 24}, 449 (1976).\\ 
M.K. Prasad and C.M. Sommerfield, Phys. Rev. Lett. {\bf 35}, 760 (1975).
     
\bibitem{Losev}
X. Hou, A. Losev, and M. Shifman,  Phys. Rev. {\bf D}{\bf 61}, (2000) 085005.\\   
G. Dvali and M. Shifman,  Phys. Lett. {\bf B454}, (1999) 277.\\ 
R. Hofmann, Phys. Rev. {\bf D62}, 065012 (2000), hep-th/0004178.

\bibitem{Parry0}
A. Sornborger and M. Parry, Phys. Rev. Lett. {\bf 83}, 666 (1999), hep-ph/9811520.\\  
A. Sornborger and M. Parry, Phys. Rev. {\bf D62}, 083511 (2000), hep-ph/0004230.
     
\bibitem{Bellido}
J.G. Garcia-Bellido and E. Ruiz Morales, hep-ph/0109230.

\bibitem{KofmanLinde}
G.N. Felder, L. Kofman, A.D. Linde, Phys. Rev. {\bf D64}, 123517 (2001), hep-th/0106179. 

\bibitem{Parry1}
M.F. Parry and A.T. Sornborger, Phys. Rev. {\bf D60}, 103504 (1999), hep-ph/9805211.

\bibitem{hof5}
R. Hofmann, hep-ph/0201073. 

\bibitem{Lyth}
D.H. Lyth and D. Wands,  Phys. Lett. {\bf B524}, (2002) 5, hep-ph/0110002.     

\bibitem{Wilczek}
L.M. Krauss and F. Wilczek, Phys. Rev. Lett. {\bf 62}, 1221 (1989).

\bibitem{Cohn}
K. Hamaguchi, Y. Nomura, and T. Yanagida, Phys. Rev. {\bf D58}, 103503 (1998), hep-ph/9805346.\\ 
K. Hamaguchi and Y. Nomura, Phys. Rev. {\bf D59}, 063507 (1999), hep-ph/9809426.


\end{thebibliography}

%

\end{document}